\documentclass{PoS}
\usepackage{amsmath,booktabs,ulem}
\usepackage{mathtools}
\usepackage[numbers,sort&compress]{natbib}
\usepackage{natbibspacing}
\usepackage{lineno}
\definecolor{darkgreen}{rgb}{0,.6,0}
\DeclareSymbolFont{greek}{OML}{cmr}{m}{n}
\DeclareMathSymbol{\epsilon}{0}{greek}{"0F}

\title{Hadronic form factors for rare semileptonic $B$~decays}

\ShortTitle{Hadronic form factors for rare semileptonic $B$ decays}

\author{Jonathan Flynn$^a$, Andreas J\"uttner$^a$, Taichi Kawanai$^b$, \speaker{Edwin
Lizarazo}$^a$,\newline Oliver Witzel$^c$ (RBC and UKQCD collaborations)\\
        \llap{$^a$} Physics and Astronomy, University of Southampton, Southampton SO17 1BJ, UK \\
        \llap{$^b$} Forschungszentrum J\"ulich, Institute for Advanced Simulation,\newline J\"ulich Supercomputing Centre, 52425 J\"ulich, Germany\\
        \llap{$^c$} Higgs Centre for Theoretical Physics, School of Physics and Astronomy,\newline The University of Edinburgh, EH9 3FD, UK \\
        E-mail: \email{j.m.flynn@soton.ac.uk, juettner@soton.ac.uk,
t.kawanai@fz-juelich.de, e.lizarazo@soton.ac.uk, o.witzel@ed.ac.uk}}

\abstract{We discuss first results for the computation of short distance 
contributions to semileptonic form factors for the rare $B$ decays $B \to K^{*} \ell^+
\ell^-$ and $B_s \to \phi \ell^+ \ell^-$. Our simulations are based on RBC/UKQCD's 
$N_f=2+1$ ensembles with domain wall light quarks and the Iwasaki gauge action. For 
the valence $b$-quark we chose the relativistic heavy quark action. }

\FullConference{The 33rd International Symposium on Lattice Field Theory\\
		14 -18 July 2015\\
		Kobe International Conference Center, Kobe, Japan}

\begin{document}

\section{Introduction}
Flavour physics plays a crucial role in the quest for new physics beyond the Standard Model (SM).
Here we concentrate on flavour changing processes involving $b$-quarks. Its large
mass may give rise to (virtual) new heavy 
particles or exhibit large enough couplings to e.g.~charged Higgs bosons 
postulated by the 2-Higgs doublet model \cite{Lee:1973iz,Donoghue:1978cj}. In order to detect 
such signs of new physics, it is important to improve our understanding 
of flavour changing processes which occur at tree and loop level in the SM.
Compared to tree level flavour changing processes in the SM, 
transitions proceeding via Flavour Changing Neutral Currents (FCNC) are highly suppressed. 
FCNC are therefore particularly attractive probes for discovering potential 
contributions from physics beyond the SM (BSM).
So far however no discovery has been made. Both experiment
and theory are working on reducing uncertainties and improving bounds. 
On the theoretical side predictions of hadronic uncertainties
 are typically dominating the error budget. Here we are presenting our efforts in 
using simulations of lattice QCD for making reliable and precise predictions 
for hadronic form factors that are crucial inputs to the analysis.

FCNC are conventionally described in the SM by a set of 20 operators 
which contribute to the effective Hamiltonian \cite{Grinstein:1987vj,
Grinstein:1990tj,Buras:1993xp,Ciuchini:1993ks,Ciuchini:1993fk,Ciuchini:1994xa}.
Weak $b\to q$ decays (with $q$ a down or a strange quark) are parametrised by
\begin{linenomath*}
\begin{equation}
\mathcal{H}^{b\to q}_\text{eff}  = -\frac{4G_F}{\sqrt{2}}V^*_{tq}V_{tb}
\sum_{i=1}^{10}
C_i O_i + C^\prime_i O^\prime_i,
\end{equation}
\end{linenomath*}
where  $V_{tq}^*$ and $V_{tb}$ are CKM matrix elements, $O_i^{(\prime)}$ are local operators and $C_i^{(\prime)}$ their 
corresponding Wilson coefficients determined in \cite{Buras:2002tp,Gambino:2003zm,Altmannshofer:2008dz}. 
The leading short distance contributions are given by
\begin{linenomath*}
\begin{align}
  O^{(\prime)}_9 &= \frac{e^2}{16\pi^2} \bar{q}\gamma^\mu P_{L(R)} b \bar{\ell}\gamma_\mu \ell,&
  &O^{(\prime)}_{10} = \frac{e^2}{16\pi^2} \bar{q}\gamma^\mu P_{L(R)} b \bar{\ell}\gamma_\mu\ell,
\label{eq:tf-2}
\end{align}
\end{linenomath*}
and the dileptonic operator
\begin{linenomath*}
\begin{equation}
	O^{(\prime)}_7 = \frac{m_be}{16\pi^2}\bar{q}\sigma^{\mu\nu}P_{R(L)} b F_{\mu\nu}. 
\label{eq:tf-1}
\end{equation}
\end{linenomath*}
In Equations (\ref{eq:tf-2}) and (\ref{eq:tf-1}) the lepton is denoted by $\ell$, the mass of the $b$-quark by $m_b$ and $P_{L(R)} = \frac{1}{2}(1\mp\gamma^5)$,
$\sigma^{\mu\nu}=\frac{i}{2}[\gamma^\mu, \gamma^\nu]$, $F^{\mu\nu} = \partial^\mu A^\nu - \partial^\nu A^\mu$.
Further, long distance effects contribute which are commonly estimated perturbatively \cite{Grinstein:2004vb, Beylich:2011aq}. These theoretical estimates are however questioned because charm resonances may not be captured 
well enough \cite{Lyon:2014hpa}. More research is needed to understand the implications and to also 
explore current tensions between SM predictions and experimental 
results. Of particular interest is e.g.~the observed tension for the observable 
$P_5^\prime$ in $B_s^∗ \to K \ell^+\ell^-$ decays \cite{Aaij:2015esa, LHCb:2015dla} 
which can also be explained by different models of new physics, see e.g.~\cite{Descotes-Genon:2013wba, Altmannshofer:2013foa, Altmannshofer:2014rta}.

In the following we focus on the computation of the dominant short distance
operators for which a lattice calculation is suitable. Moreover, in these proceedings we restrict ourselves to the computation of decays with a pseudoscalar $B_{(s)}$ meson in the initial and a vector meson in the final state. 
As in all other contemporary computations of these decays the final state vector meson will be 
approximated as a stable particle. Our computations provide important, independent checks to existing calculations by the Cambridge group \cite{Horgan:2013pva,Horgan:2013hoa,Horgan:2015vla}, HPQCD \cite{Bouchard:2013mia}, and the Fermilab/MILC \cite{Bailey:2015dka,Du:2015tda} collaborations which are all based on overlapping sets of MILC's staggered gauge field configurations. In the following Section we describe our computational setup to obtain these contributions and present in Sec.~\ref{Sec.Results} our first numerical results before summarising in Sec.~\ref{Sec.Summary}.

\section{Computational setup}\label{Sec.Setup}
Our simulations are performed using 2+1 flavour domain-wall fermions and Iwasaki gauge-field ensembles generated by the RBC and UKQCD Collaborations \cite{Allton:2008pn, Aoki:2010dy}. In the valence sector we use the domain-wall action \cite{Shamir:1993zy,Furman:1994ky} for $u/d$ and $s$-quarks and the relativistic heavy quark action \cite{Christ:2006us} with nonperturbatively tuned parameters \cite{Aoki:2012xaa} for the $b$-quarks. Further details of our simulations are summarised in Tab.~\ref{tab:1}. Autocorrelations between our lattices are effectively reduced by shifting the gauge fields by a random 4-vector as first step of our simulations.

\begin{table}
\label{tab:1}
\centering
\begin{tabular}{cccccccc}
\toprule
$L^3 \times T$ & $a^{-1}$ [GeV] & $am_l$ & $am_h$ & $am^\prime_s$  & $M_\pi$ [MeV]  & total \# of configs\\ \midrule
$24^3 \times 64$ &1.785(5)    & 0.005 & 0.040 & 0.0343 & 338 &1636  \\ 
$24^3 \times 64$ &1.785(5)    & 0.010 & 0.040  & 0.0343 & 434 &1419  \\ 
\bottomrule
\end{tabular}
\caption{Parameters of the calculation. We use the coarse ensembles generated by the RBC and UKQCD collaboration \cite{Allton:2008pn, Aoki:2010dy} with 2+1 flavor domain-wall fermions and Iwasaki gauge actions. The domain-wall fermions are created with fifth dimension $L_s=16$ and domain-wall height $M_5=1.8$. Values for the inverse lattice spacing
and the quark and meson masses are taken from the refined analysis \cite{Blum:2014tka}; $am_l$ labels the mass of the light sea-quark, $am_h$ the mass of the heavy sea quark, and $am_s^\prime$ the mass of the near physical strange quark mass. The physical value of the strange quark mass is $am_s=0.03224(18)$ \cite{Blum:2014tka}.}
\end{table}

We parametrise the contributions of the short distance operators defined in
Eqs.~(\ref{eq:tf-2}) and (\ref{eq:tf-1}) by a set of seven form factors,
$f_V,\, f_{A_0},\, f_{A_1},\, f_{A_2},\, f_{T_1},\, f_{T_2},$ and $f_{T_3}$
assuming we have a $B_{(s)}$ meson in the initial and a light vector meson
($V$) in the final state.  We determine the seven form factors by evaluating
the following hadronic weak matrix elements \cite{Altmannshofer:2008dz}: 
\begin{linenomath*}
\begin{align}
\langle V( k, \varepsilon) | \bar{q} \gamma^\mu b | {B_{(s)}}(p) \rangle 
   &= {f_V(q^2)} \frac{2i\epsilon^{\mu\nu\rho\sigma} 
		\varepsilon_{\nu}^*  k_\rho p_\sigma}{M_{B_{(s)}} + M_V} \label{eq:1}\\
\langle V( k, \varepsilon) | \bar{q} \gamma^\mu \gamma_5 b | {B_{(s)}}(p) \rangle 
   &= {f_{A_0}(q^2)}\frac{2M_V \varepsilon^*\cdot q}{q^2}   q^\mu \notag\\
     &\quad + {f_{A_1} (q^2)}(M_{B_{(s)}} + M_V)\left[ \varepsilon^{*\mu}  -
\frac{\varepsilon^*\cdot q}{q^2}   q^\mu \right] \notag\\
    &\quad - {f_{A_2}(q^2)} \frac{\varepsilon^*\cdot q}{M_{B_{(s)}}+M_V} \left[
k^\mu + p^\mu - \frac{M_{B_{(s)}}^2 -M_V^2}{q^2}q^\mu\right] \\ 
q_\nu	 \langle V( k, \varepsilon) | \bar{q} \sigma^{\nu\mu} b | {B_{(s)}}(p) \rangle 
   &= 2{f_{T_1}(q^2)} \epsilon^{\mu\rho\tau\sigma} \varepsilon_{\rho}^*
k_\tau p_\sigma \\
q_\nu \langle V( k, \varepsilon) | \bar{q} \sigma^{\nu\mu} \gamma^5 b | {B_{(s)}}(p) \rangle 
      &= i {f_{T_2}(q^2)} \left[
	\varepsilon^{*\mu}  (M_{B_{(s)}}^2- M_V^2) -  (\varepsilon^* \cdot q)(p + k)^\mu
      \right] \notag\\
&\quad+ i{f_{T_3} (q^2)}  (\varepsilon^* \cdot q)\left[  
	q^\mu - \frac{q^2}{M_{B_{(s)}}^2 - M_V^2} (p + k)^\mu
       \right]\label{eq:2}
\end{align}
\end{linenomath*}
where $p$ ($k$) is the momentum of the $B_{(s)}$ (vector) meson and 
$q = (M_{B_{(s)}}-E_V(|\vec{k}|),-\vec{k})$. 
The masses of the $B_{(s)}$ and the vector meson are given by $M_{B_{(s)}}$ and $M_V$, respectively, and $\varepsilon$ denotes the polarisation vector of the final state meson.

\begin{figure}[t]
\centering
\includegraphics{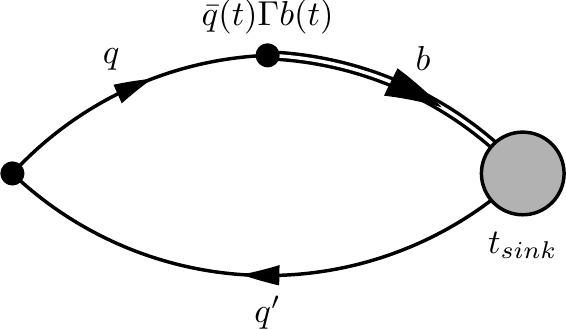}
\caption{Three point correlator function used to obtain the $B\to V$ form
factors. The single and double lines correspond to light and $b$-quark
propagators, respectively.} 
\label{fig:3pt}
\end{figure}

The matrix elements Eqs.~(\ref{eq:1}) -- (\ref{eq:2}) are evaluated by computing ratios of 3-point over 2-point correlation functions and we choose to carry out the computation in the $B_{(s)}$-meson rest frame 
\begin{linenomath*}
\begin{align}\label{eq:ratio}
 R_{B_{(s)} \to V}^{\alpha\Gamma} (t,t_\text{sink},{k}) 
&=  \frac{C_{B_{(s)}\to V}^{\alpha\Gamma}(t,t_\text{sink},{k}) }
{\sqrt{\frac{1}{3}\sum_i C^{ii}_{V}(t,{k}) C_{B_{(s)}} (t_\text{sink}-t)}}
\sqrt{\frac{4E_V M_{B_{(s)}}\sum_\lambda \varepsilon^j(k,\lambda)\varepsilon^{j*}(k,\lambda)}
{e^{-E_V t} e^{-M_{B_{(s)}}(t_\text{sink}-t)}}}\\
&\xrightarrow{t,t_\text{sink} \to \infty}
\sum_\lambda  \varepsilon^\alpha(k, \lambda) \langle V(k, \lambda) 
|  \bar{q}\Gamma b| B_{(s)}(p) \rangle
\end{align}
\end{linenomath*}
The 3-point functions
\begin{linenomath*}
\begin{equation}
C^{\alpha\Gamma}_{B_{(s)}\to V}(t, t_\text{sink}, \vec{k}) = 
\sum_{\vec{x},\vec{y}}e^{i\vec{k}\cdot\vec{y}}\langle
\mathcal{O}^\alpha_V(0,\vec{0})\bar{q}(t,\vec{y})\Gamma b(t,\vec{y})\mathcal{O}_{B_{(s)}}(t_\text{sink},\vec{x})\rangle ,
\label{eq.3pt}
\end{equation}
\end{linenomath*}
are sketched in Fig.~\ref{fig:3pt} and constructed in the following way: 
A light quark propagator with the appropriate Dirac-structure and Fourier phase at its 
sink is used as the source for a sequential $b$-quark inversion which is then contracted 
with another light quark propagator originating form the same
location as the first one. The contractions are performed by inserting a vector or 
tensor current
$\bar{q}\Gamma b$ with 
$\Gamma = \{\gamma^\mu, \gamma^5\gamma^\mu, \sigma^{\mu\nu}, \gamma^5\sigma^{\mu\nu}\}$ 
between the interpolating meson operators $\mathcal{O}_V = \bar{q}\gamma^\mu q$ and
$\mathcal{O}_{B_{(s)}} = \bar{b}\gamma^5q$.  As usual
we project the result onto
states of discrete momenta $\vec k$.  
The 2-point functions in the denominator of Eq.~(\ref{eq:ratio}) are given by
\begin{linenomath*}
\begin{align}
C_{B_{(s)}}(t) &= \sum_{\vec{x}}\langle
\mathcal{O}^\dagger_{B_{(s)}}(t,\vec{x})\mathcal{O}_{{B_{(s)}}}(0,\vec{0})\rangle,&
\text{and}&
&C_V(t,\vec{k}) = \sum_{\vec{x}}e^{i\vec{k}\cdot \vec{x}}\langle
\mathcal{O}^\dagger_V(t,\vec{x})\mathcal{O}_{{ V}}(0,\vec{0})\rangle.
\label{eq:2pt-0}
\end{align}
\end{linenomath*}
In addition we use the correlators \eqref{eq:2pt-0} to extract the $B_{(s)}$ meson mass, $M_{B_{(s)}}$, and the energy of the vector meson $V$ with momentum $\vec k$, $E_V(|\vec k|)$.

We reduce excited state contamination by generating the heavy $b$-quark propagators 
with a Gaussian smeared source using the parameters determined in
 \cite{Aoki:2012xaa}, while light $u/d$ and $s$-quark propagators are generated from a point source.

\section{First results}\label{Sec.Results}
\begin{figure}[t]
\centering
\includegraphics[scale=0.3]{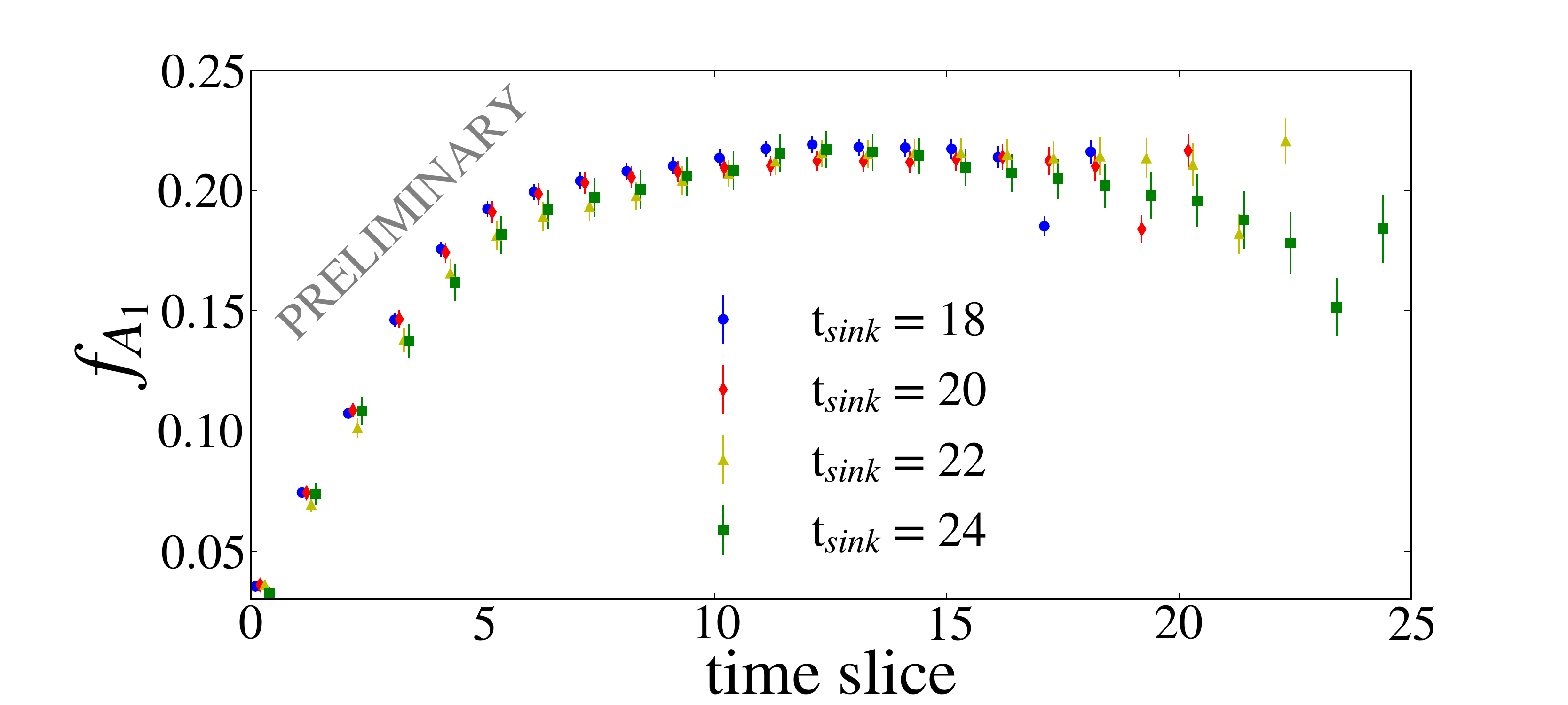}
\caption{Bare form factor $f_{A_1}$ for the $B_s\to\phi\ell^+\ell^-$
decay at zero momentum for four different source-sink separations $t_{sink}$ on the 
coarse $a\approx0.11$fm ensemble with $am_l=0.005$.}
\label{fig:source-sink}
\end{figure}

\begin{figure}[t]
\centering
\parbox{0.495\textwidth}{\includegraphics[width=\linewidth]{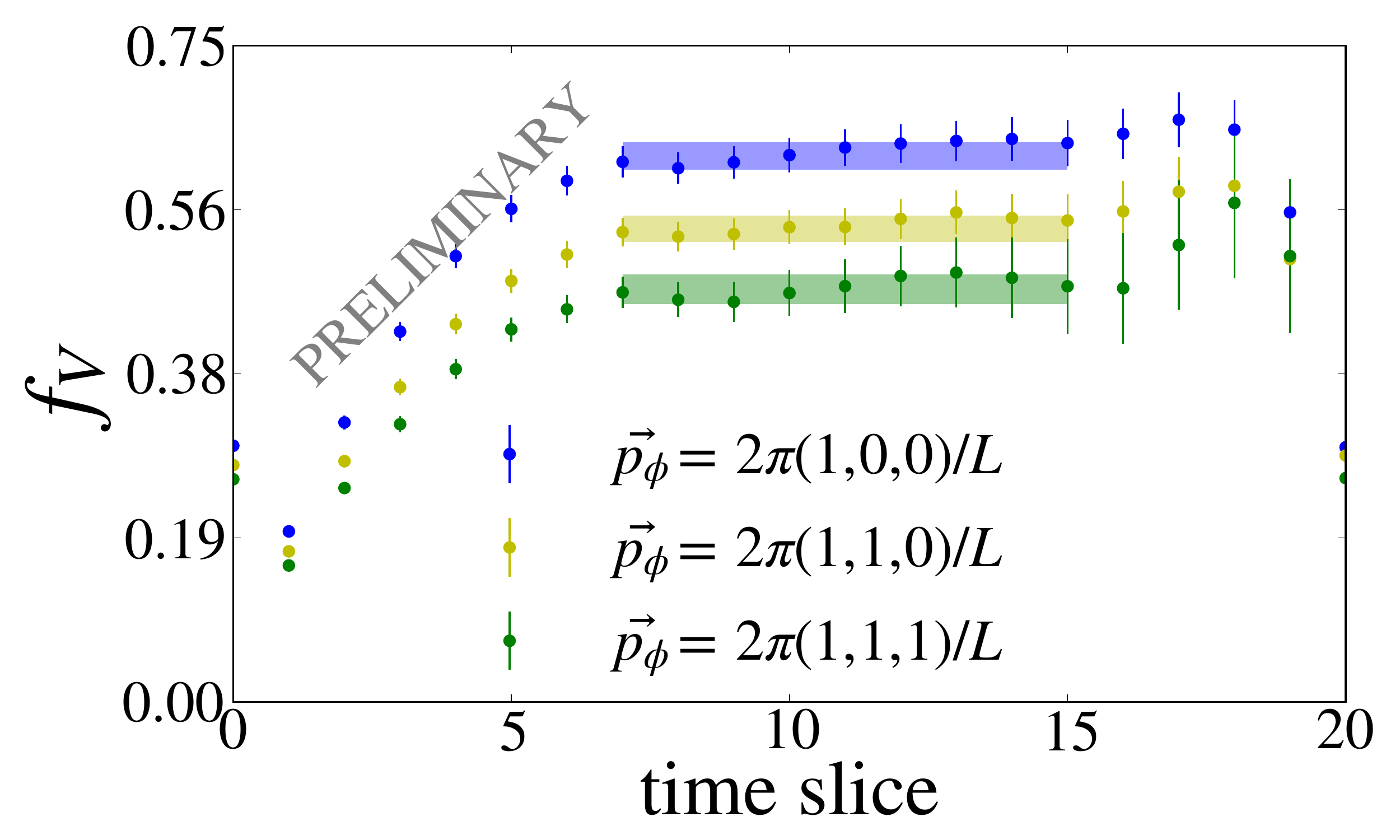}}
\parbox{0.495\textwidth}{\includegraphics[width=\linewidth]{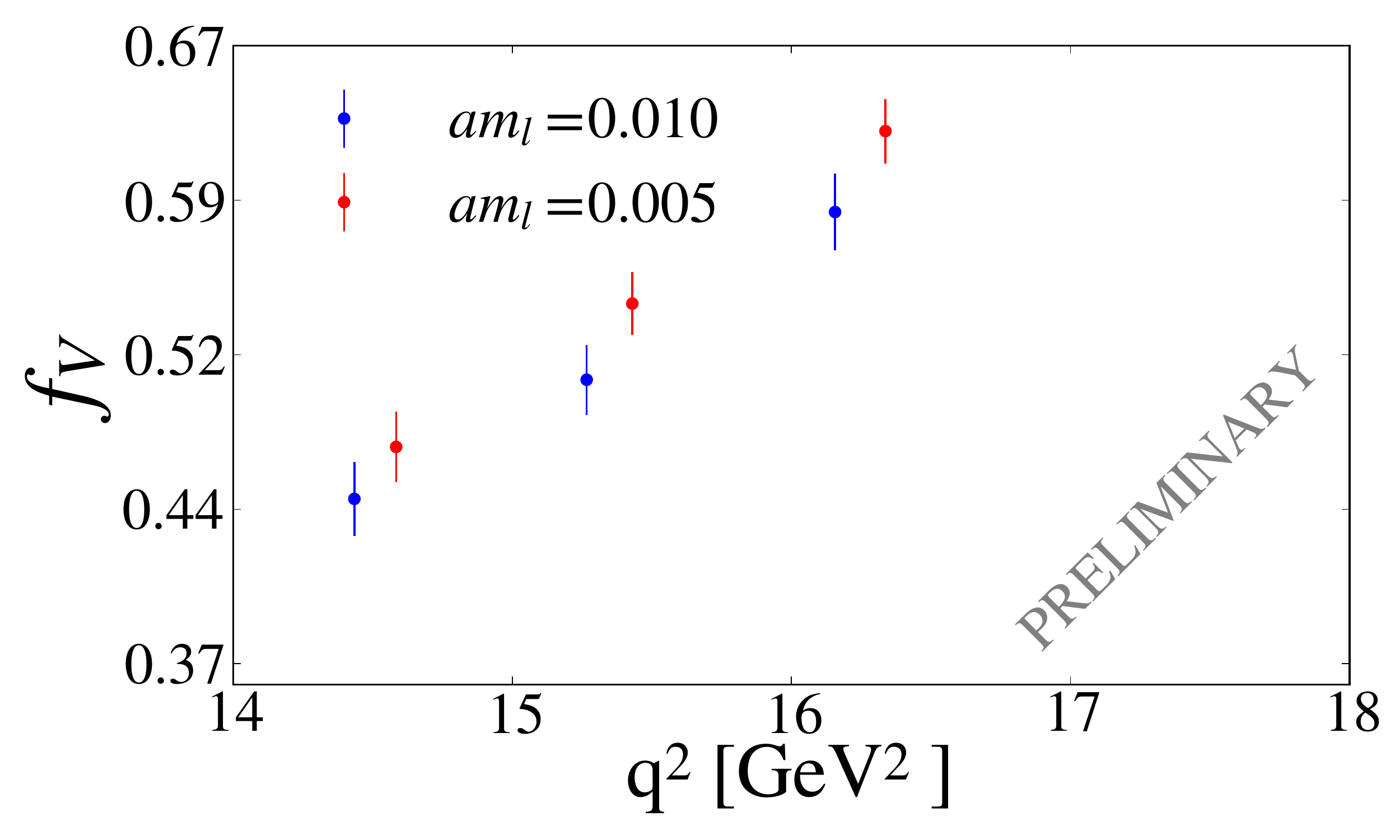}}
\parbox{0.495\textwidth}{\includegraphics[width=\linewidth]{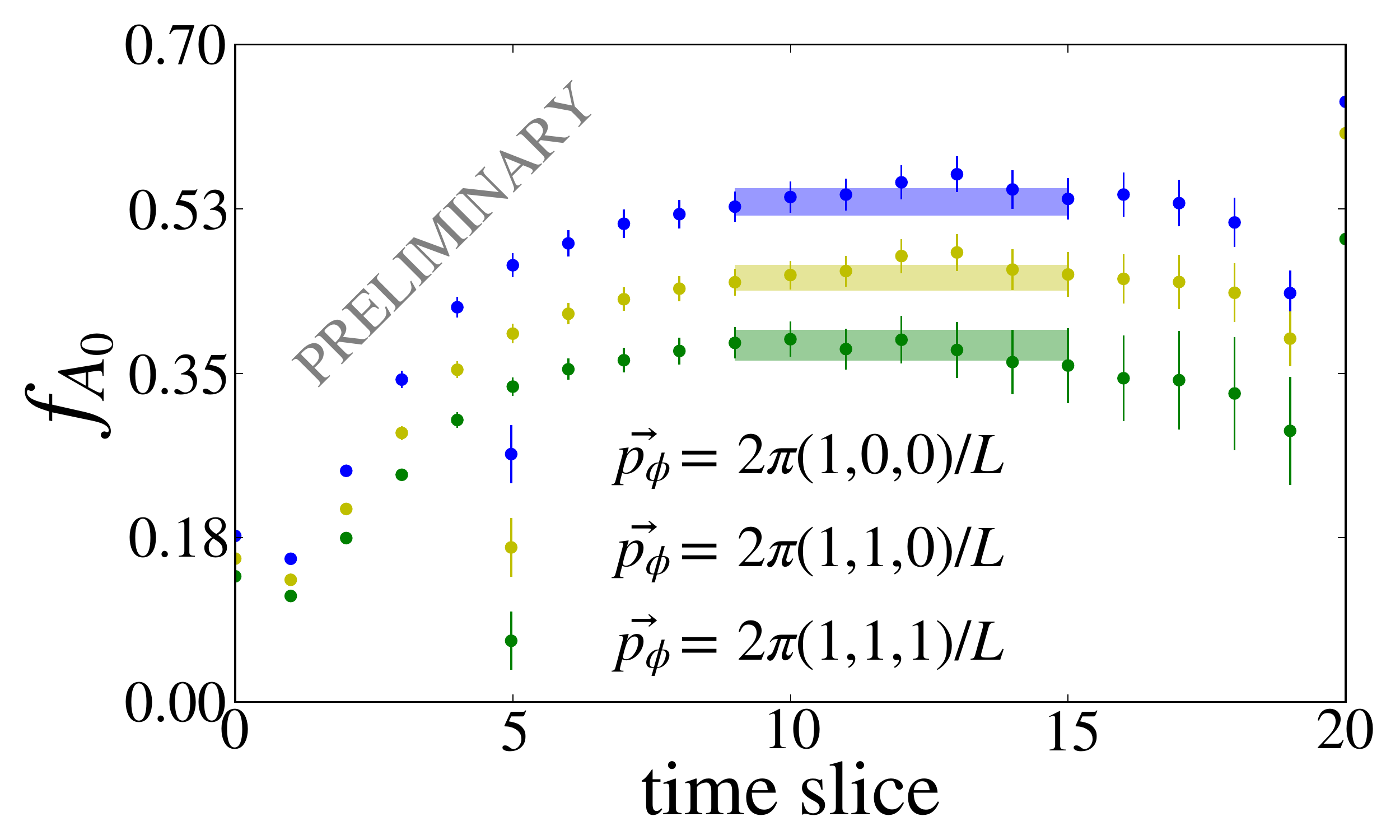}}
\parbox{0.495\textwidth}{\includegraphics[width=\linewidth]{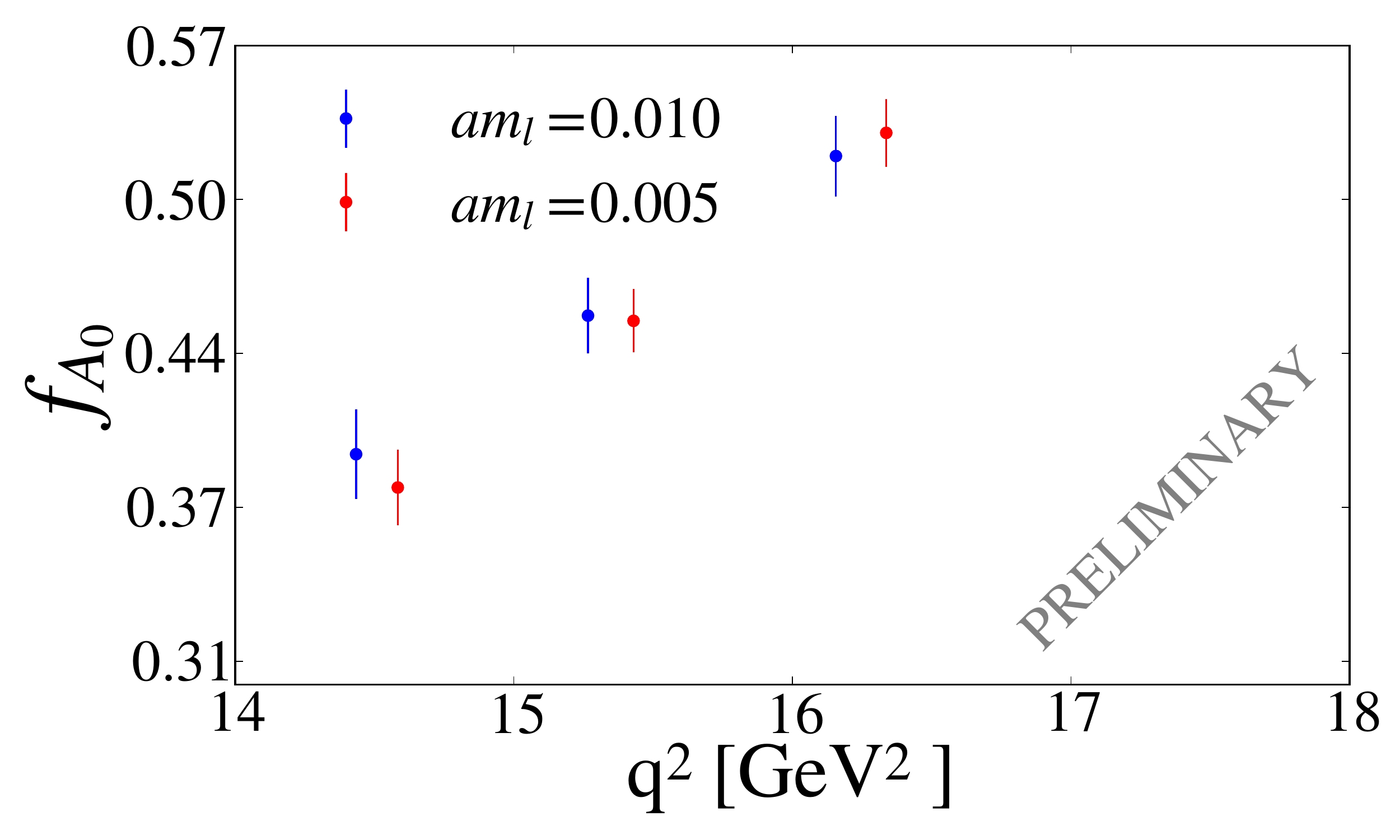}}
\parbox{0.495\textwidth}{\includegraphics[width=\linewidth]{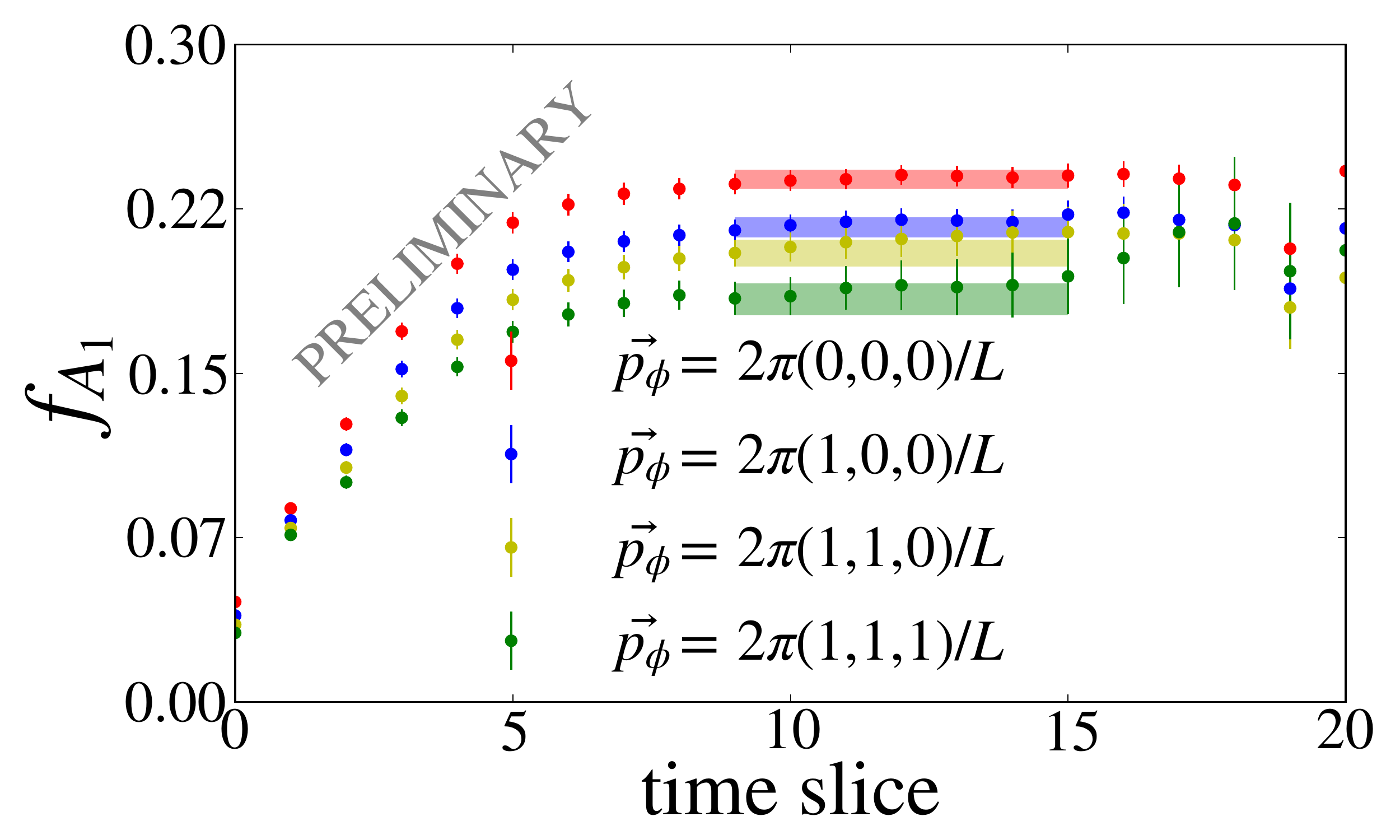}}
\parbox{0.495\textwidth}{\includegraphics[width=\linewidth]{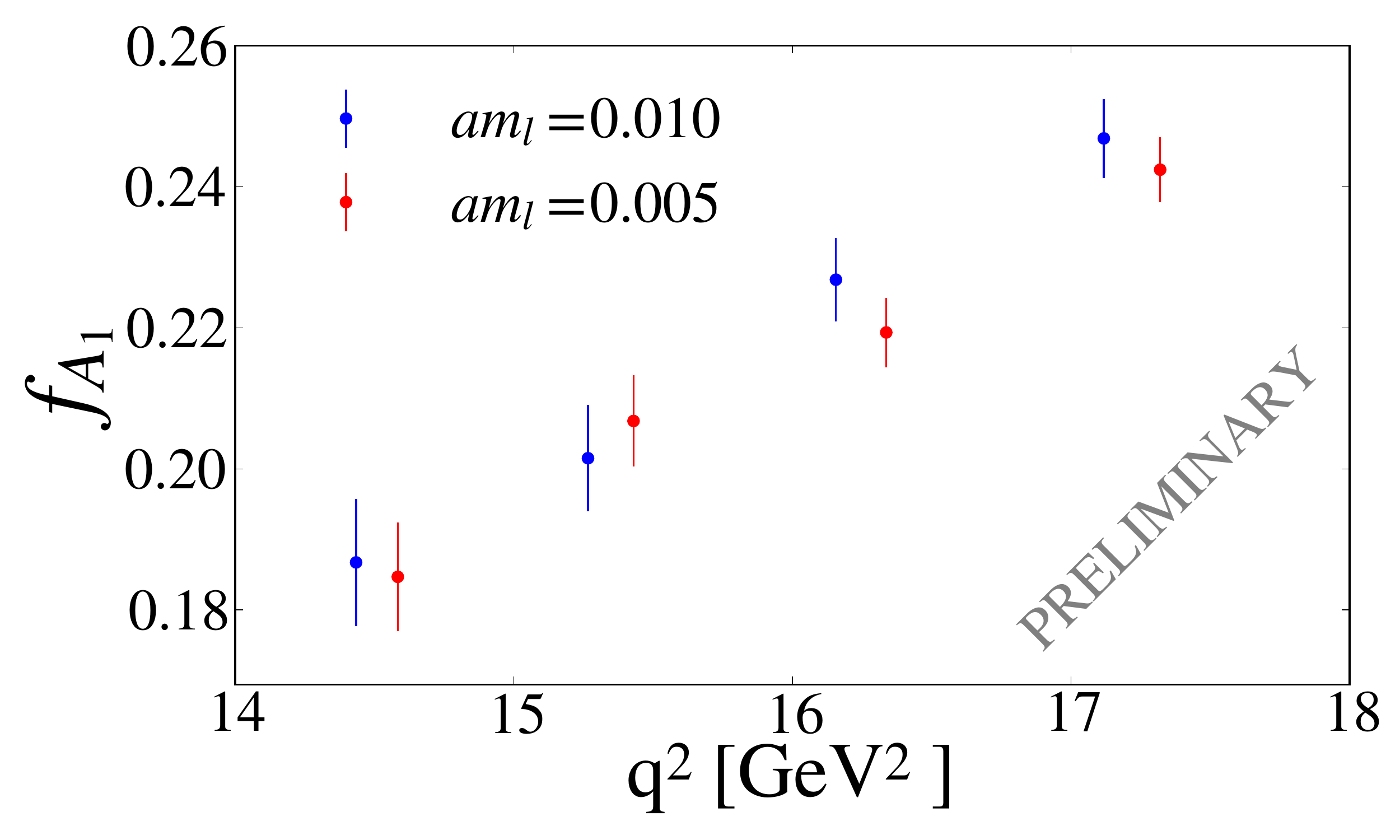}}
\parbox{0.495\textwidth}{\includegraphics[width=\linewidth]{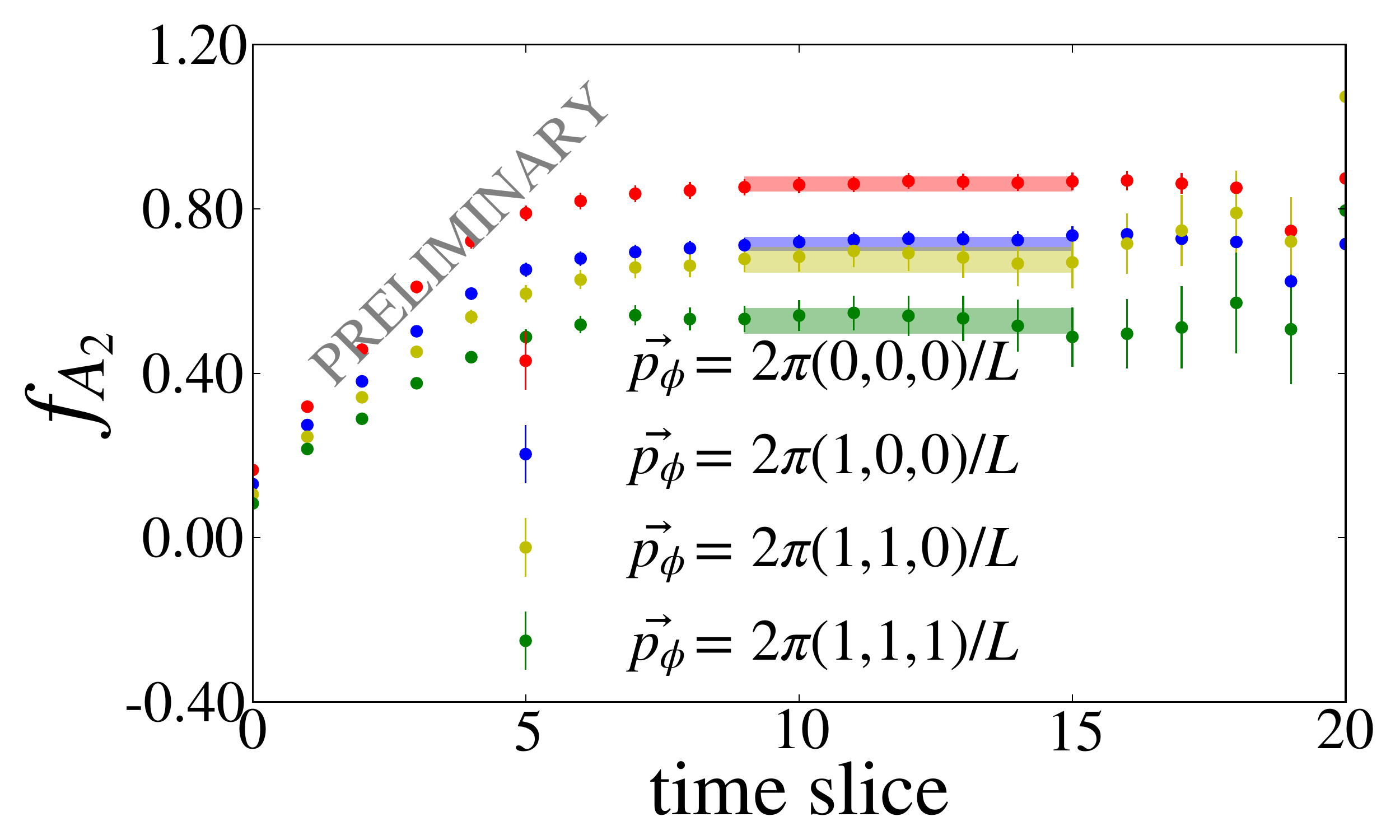}}
\parbox{0.495\textwidth}{\includegraphics[width=\linewidth]{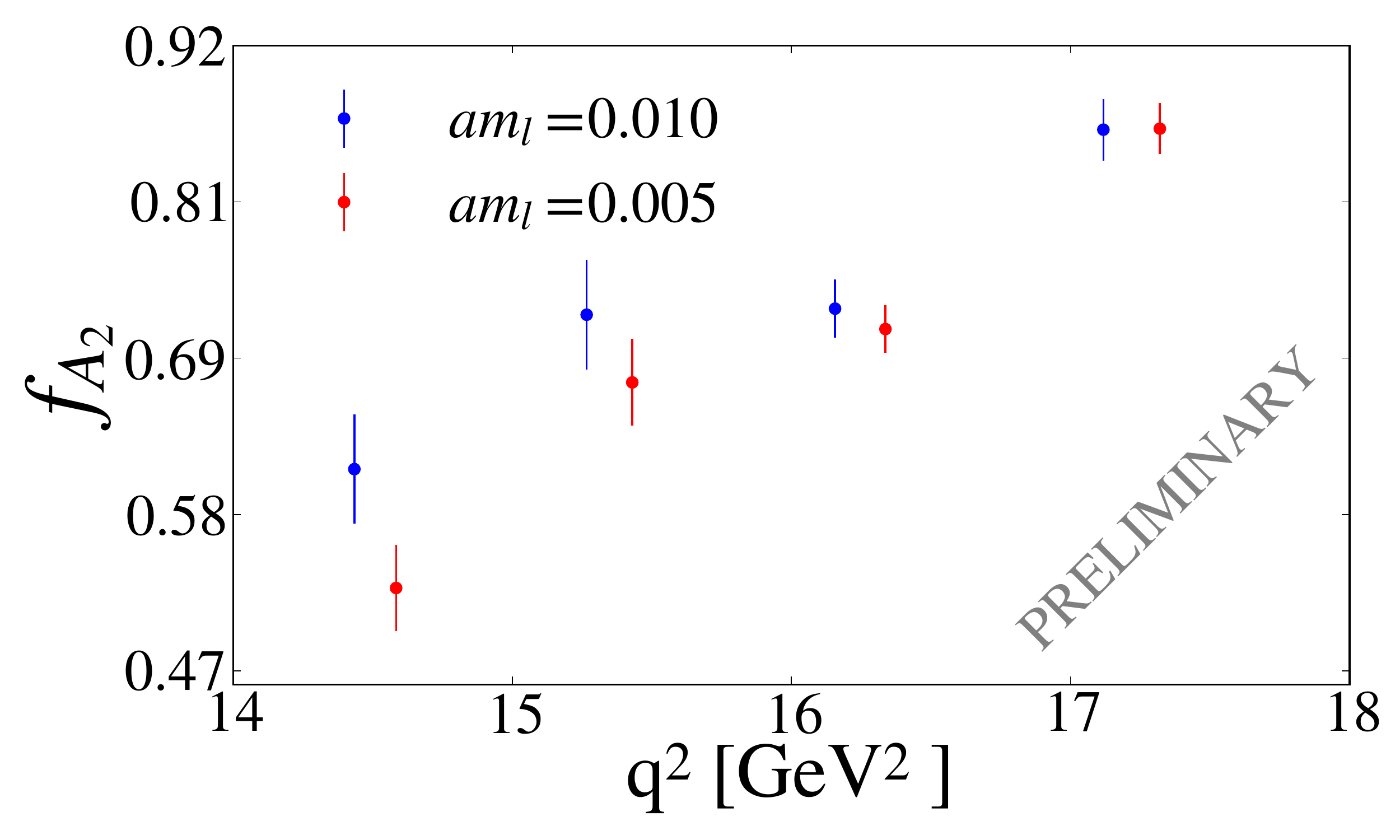}}
\caption{Bare form factors $f_V,\ f_{A_0},\ f_{A_1},\ f_{A_2}$ for $B_s \to \phi \ell^+\ell^-$
decays. Left: plateau fits for the coarse ensemble with $am_l=0.005$, right:
extracted form factor values vs.~$q^2$ in GeV obtained on both coarse
ensembles, $am_l = 0.005$ and $am_l = 0.010$.}
\label{fig:A1}
\end{figure}

Central to the numerical part of this project is the calculation of the 3-point functions defined in Eq.~(\ref{eq.3pt}). We implement those as inline function in Chroma \cite{Edwards:2004sx}. In order to optimise the signal, we study four different source-sink
separations seeking the choice which results in the longest plateau and
small statistical errors.  The outcome is shown in
Fig.~\ref{fig:source-sink} where we compare the form factor $f_{A_1}$ 
for the $B_s\to\phi\ell^+\ell^-$ decay at zero momentum using $t_\text{sink}=18,\,20,\,22,\,24$. Within statistical uncertainties, all choices for $t_\text{sink}$ agree. The best signal is however obtained for $t_\text{sink}=20$. We extract the value of the 3-point functions by fitting the range of the plateau sufficiently far from the source and sink locations in order to minimise the influence of excited states. 
In this way we have obtained results for all form factors $f_V,\, f_{A_0},\, f_{A_1}$, $f_{A_2}$, $f_{T_1},\, f_{T_2}$, and $f_{T_3}$, some of which we show in Fig.~\ref{fig:A1}. For all form factors we find good signals and can extract values from correlated fits to plateaus ranging over several time slices.

\section{Summary and outlook}\label{Sec.Summary}
In this paper we present our first results computing the short distance contributions to $B_s\to \phi\ell^+\ell^-$ decays on the lattice. We have validated our code and calculated the form factors parametrising this rare FCNC decay using two ensembles at our coarse lattice spacing of $a^{-1}=1.785$ GeV. In parallel we are computing renormalisation factors using a {\it mostly nonperturbative setup}\/ \cite{ElKhadra:2001rv} and are also analysing form factors for $B\to K^* \ell^+\ell^-$ decays.

The next steps are to include additional terms for the 1-loop $O(a)$ improvement of the operators and subsequently recalculate the form factors on the two coarse ensembles presented here and in addition on three finer ensembles with $a^{-1}=2.38$ GeV. Moreover we are exploring to add ensembles with physical light quark masses or an even finer lattice spacing ($a^{-1}\approx 2.8$ GeV). With a set of measurements obtained for different sea- and valence-quark masses and at different values of the lattice spacing, we then intend to perform a combined chiral- and continuum extrapolation based on heavy meson chiral perturbation theory in order to facilitate a kinematical extrapolation ($z$-expansion) to zero $q^2$.

\clearpage
\paragraph{Acknowledgements}
The authors thank our collaborators in the RBC and UKQCD Collaborations
for helpful discussions and suggestions. Computations for this work 
were performed on resources provided by the USQCD Collaboration, funded by 
the Office of Science of the U.S. Department of Energy, as well as on 
computers at Columbia University and Brookhaven National Laboratory. Gauge 
field configurations on which our calculations are based were also generated using the
DiRAC Blue Gene Q system at the University of Edinburgh, part of the DiRAC
Facility; funded by BIS National E-infrastructure grant ST/K000411/1 and 
STFC grants ST/H008845/1, ST/K005804/1 and ST/K005790/1.  The
research leading to these results has received funding from the European
Research Council under the European Unions Seventh Framework Programme 
(FP7/2007-2013) / ERC Grant agreement 279757, STFC grant ST/L000296/1 and ST/L000458/1 
as well as the EPSRC Doctoral Training Centre grant (EP/G03690X/1).

{\small
\bibliography{B_meson}
\bibliographystyle{apsrev4-1ow}
}



\end{document}